\documentclass[reprint,superscriptaddress,showpacs,amsmath,amssymb,nofootinbib,aps]{revtex4-1}
\usepackage{graphicx}
\usepackage{dcolumn}
\usepackage{bm}
\usepackage{amsthm,amsmath,amssymb}
\usepackage{mathrsfs}
\usepackage{multirow}
\usepackage[T1]{fontenc}
\usepackage{blindtext}
\usepackage{xcolor}
\usepackage[
colorlinks=true,        
citecolor=blue,         
linkcolor=blue,         
urlcolor=blue           
]{hyperref}             
\usepackage{float}

\begin{document}
\preprint{APS/123-QED}

\title{Constraints on Primordial Black Hole Dressed by Dark Matter Halo from Microlensing Effect of Fast Radio Bursts}

\author{Hong-Rui Tao}
\affiliation{School of Physics and Optoelectronic Engineering, Yangtze University, Jingzhou, 434023, China}

\author{Huan Zhou}
\email{huanzhou@yangtzeu.edu.cn \textcolor{black}{(Corresponding Author)}}
\affiliation{School of Physics and Optoelectronic Engineering, Yangtze University, Jingzhou, 434023, China}

\author{Cheng-Gang Shao}
\email{cgshao@hust.edu.cn}
\affiliation{School of Physics and Optoelectronic Engineering, Yangtze University, Jingzhou, 434023, China}

\author{Xiao-Long Gong}
\affiliation{School of Physics and Optoelectronic Engineering, Yangtze University, Jingzhou, 434023, China}

\author{Zheng-Xiang Li}
\affiliation{School of Physics and Astronomy, Beijing Normal University, Beijing 100875, China}
\affiliation{Institute for Frontiers in Astronomy and Astrophysics, Beijing Normal University, Beijing 102206, China}
\date{\today}

\begin{abstract}
Primordial black holes (PBHs) are not only considered as a candidate for dark matter, but also as potential sources of gravitational waves from binary black hole mergers by the LIGO-Virgo-KAGRA and as seeds for the supermassive black holes observed by the James-Webb Space Telescope, thereby remaining intense interest in cosmology and astrophysics. Fast radio bursts (FRBs) are bright millisecond-duration radio transients whose physical origin remains elusive, which have rapidly developed into one of the most active and rapidly evolving fields in astronomy. The microlensing effect of FRBs offers a clean and powerful probe of PBHs, especially in the mass range above stellar-mass window. In this work, we derive a complete transformation that converts any upper limit on the abundance of PBHs originally derived for `bare' PBHs with monochromatic mass distribution, into the corresponding constraint on `dressed' PBHs with arbitrary extended mass distributions. Based on this framework, we estimate the future constraints on the dressed PBH abundance \(f_{\mathrm{PBH}}\) from FRB observations assuming an expected sample of \(10^5\) FRBs accumulated over the next decade well within the projected detection capabilities of SKA. Our results indicate that including halo enhancement tightens the upper limits on \(f_{\mathrm{PBH}}\) by approximately one order of magnitude, with the most stringent constraint reaching \(\sim10^{-4}\) for the typical mass range from stellar-mass to intermediate-mass black holes.
\end{abstract}
\maketitle

\section{Introduction}
Primordial black holes (PBHs)~\cite{Hawking:1971ei,Carr:1974nx,Carr:1975qj}, which formed in the early universe and are considered as a potential component of dark matter, exhibit a wide range of mass windows-spanning from the Planck mass scale ($10^{-5}~\rm g$) to supermassive black holes (SMBHs) in galactic centers. Recent developments in multi‑wavelength and multi‑messenger astronomy have transcended the view of PBHs as merely dark matter candidates and have positioned them as a key theoretical framework for interpreting a wide range of astrophysical phenomena. In gravitational-wave astronomy, PBHs provide a viable channel for interpreting the binary black hole merger data reported by the LIGO-Virgo-KAGRA collaborations~\cite{Sasaki:2016jop,Ali-Haimoud:2017rtz,Chen:2018czv}. In particular, they provide an interpretation for specific binary black hole merger events~\cite{DeLuca:2025fln}, e.g., intermediate-mass black hole mergers~\cite{LIGOScientific:2020iuh,LIGOScientific:2025rsn} and sub-solar-mass merger events~\cite{LIGOScientific:2024elc,Haque:2026yum}. In the field of galaxy cosmology, James Webb Space Telescope (JWST) observations have revealed numerous quasars powered by SMBHs existing as early as the first few hundred million years after the Big Bang~\cite{CEERSTeam:2023qgy,Goulding:2023gqa,Maiolino:2023bpi,Maiolino:2023zdu, Bogdan:2023ilu, Natarajan:2023rxq, Kovacs:2024zfh, Maiolino:2025tih,Juodzbalis2025}. The presence of these SMBHs at high redshifts, particularly their overmassive nature relative to their host galaxies, represents a major challenge to the standard $\Lambda$CDM model. A prominent example is the system Abell2744-QSO1, which hosts an SMBH with mass of $\sim5\times10^7~M_{\odot}$, extreme black-hole-to-stellar-mass ratio of $\geq2$, and low metallicity of $\leq0.01~Z_{\odot}$, a combination that contradicts conventional formation models~\cite{Maiolino:2025tih, Juodzbalis2025}. These observations provide strong motivation for the early seeding scenario involving PBHs, which could naturally account for such anomalous systems~\cite{Dayal:2025aiv, Zhang:2025oyl, DeLuca:2025nao}. Over the past several decades, extensive observational searches for PBHs have yielded a variety of methods to constrain the PBH abundance across different mass windows, where the abundance is commonly quantified as the fraction of dark matter in PBHs, $f_{\mathrm{PBH}}\equiv \Omega_{\mathrm{PBH}} / \Omega_{\mathrm{DM}}$~\cite{Sasaki2018,Green:2020jor,Carr:2020gox,Carr:2021bzv,2026NCimRtmp4C}. Gravitational lensing is a powerful direct probe for constraining PBH abundance from planetary scales ($\sim 10^{-7}\,M_{\odot}$) to SMBHs in galactic centers ($\sim 10^{8}\,M_{\odot}$), complementing other methods such as Hawking radiation, gravitational waves, dynamical effects, accretion effects, and future multi-messenger observations~\cite{Sasaki2018,Green:2020jor,Carr:2020gox,Carr:2021bzv,2026NCimRtmp4C}. Depending on the observational targets and underlying physical processes, this approach can be broadly categorized into the following several types~\cite{Liao:2022gde}: 1) luminosity variations of stars within the Milky Way and nearby galaxies induced by micro-lensing~\cite{Niikura:2017zjd,Mroz:2024mse}; 2) time delayed multi-peak structures in transient sources (e.g., fast radio sources, gamma-ray bursts)~\cite{Munoz:2016tmg,Liao:2020wae,Nemiroff:2001bp,Ji:2018rvg}; 3) waveform distortions in transient sources (e.g., gravitational waves, fast radio sources) due to gravitational lensing~\cite{Jung:2017flg,CHIMEFRB:2022xzl,LIGOScientific:2025cwb}; and 4) angularly separated multiple images in persistent sources, (e.g., compact radio sources)~\cite{Wilkinson:2001vv,Zhou:2021tvp}. 

Fast radio bursts (FRBs), first discovered by Lorimer in 2007~\cite{Lorimer:2007qn}, are millisecond-duration, high-intensity radio pulses at cosmological distances, with isotropic-equivalent energies typically ranging from $\sim10^{36}~{\rm erg}$ to $10^{41}~{\rm erg}$~\cite{Cordes:2019cmq,Petroff:2019tty}. Only $2\%\sim3\%$ of FRBs are observed to repeat~\cite{CHIMEFRB:2023myn,TheCHIMEFRB:2026nji}, which probably represents a lower bound, and the intrinsic repeating fraction could be as high as $50\%$~\cite{Yamasaki:2023dlb}. The recent detection of a Galactic FRB in association with a soft gamma-ray repeater suggests that magnetar engines can produce at least some FRBs~\cite{Zhang:2020qgp, CHIMEFRB:2020abu, Bochenek:2020zxn, Lin:2020mpw}, the radiation mechanism and progenitors of FRBs remain intensively debated. However, FRBs exhibit several unique observational properties, i.e., their clean temporal structure, short duration, cosmological origin, and high all-sky event rate ($\sim10^3-10^4~{\rm sky^{-1}~day^{-1}}$~\cite{Cordes:2019cmq,Petroff:2019tty}), which position them as promising cosmological and astrophysical probes. Notable examples include constraining the expansion history of the universe and the epoch of reionization via the dispersion measure of FRBs~\cite{Deng:2013aga, Linder:2020aru, Beniamini:2020ane, Liu:2022bmn, Maity:2024zwa, Shaw:2024slf, Zhang:2025thh,Liu:2026epv}, constraining the photon mass and testing the equivalence principle through frequency-dependent time delays~\cite{Wei:2015hwd,Wu:2016brq,Chang:2024hnn}, measuring the intergalactic magnetic field~\cite{Ravi:2016kfj, Akahori:2016ami, Hackstein:2019abb,Hackstein:2020mxc,Khrykin:2025xwi}. Furthermore, lensing FRBs serves as a powerful astrophysical probe, including probing dark matter with strong or microlensed FRBs~\cite{Munoz:2016tmg, Sammons:2020kyk, Liao:2020wae,Krochek:2021opq,Zhou:2021ndx,Oguri:2022fir,Zhou:2026flq, Gao:2023xbi,Gao:2024upn}, and precisely measuring the cosmic expansion rate and curvature from time-delay of strongly lensed FRBs~\cite{Li:2017mek,Wucknitz:2020spz,Zhang:2024rra}. In recent years, the number of FRBs has increased rapidly, from several hundred in 2022 to several thousand in 2026~\cite{CHIMEFRB:2023myn,TheCHIMEFRB:2026nji}, owing to by the observational capabilities of new-generation radio telescopes operated by various countries, as well as by improvements in search strategies. This trend will be further strengthened by next-generation facilities (e.g., CHIME, DSA-2000, BURSTT, SKA), which will expand the FRB sample and make FRBs increasingly powerful for constraining PBHs~\cite{Munoz:2016tmg,Laha:2018zav,Oguri:2022fir,Connor:2022bwl,Santos:2026qnl}.

Although the existence of dark matter has been supported by multiple lines of evidence, its composition remains undetermined, with candidates including WIMPs, axions, and PBHs~\cite{Hui:2021tkt,OHare:2024nmr}. PBHs formed in the early Universe via spherical gravitational collapse of overdense regions, acting as seeds for dark matter halo growth~\cite{Ricotti:2007au}. Halo mass growth is negligible during radiation domination but can increase by up to two orders of magnitude after the Universe enters matter domination~\cite{Mack:2006gz}. Furthermore, if the PBH abundance exceeds a threshold \(f_{\rm PBH}>10^{-4}\)), the host halo can accelerate its formation by gravitationally attracting and merging with neighboring minihalos~\cite{Zhang:2024ytf}. Consequently, even stellar-mass PBHs that are a minor dark matter component are cloaked in massive halos of the dominant dark matter. This enhancement of gravitational lensing tightens PBH abundance constraints from stellar microlensing by orders of magnitude~\cite{Oguri:2022fir, Urrutia:2023mtk, Cai:2022kbp}. In this paper, we derive a transformation that converts upper limits on the abundance of bare PBHs with monochromatic mass distribution (MMD) into constraints on dressed PBHs with extended mass distributions (EMD). Under this framework, we estimate future constraints on the dressed PBH abundance \(f_{\mathrm{PBH}}\) with next-generation radio telescopes.

This paper is organized as follows: In Section~\ref{sec:method}, we introduce the model of dressed PBHs surrounded by dark matter and derive the key analytical expression based on optical depth theory. In Section~\ref{sub:three}, based on reduced analytical formulation, we forecast the constraints on PBH abundance with future FRB observations. Finally, Section~\ref{sub:four} presents conclusion and discussion. Throughout this paper, we adopt the concordance \(\Lambda\)CDM cosmological model with the best-fitting parameters from the latest Planck observations~\cite{Planck2018}, and use natural units with \(G=c=1\) in all equations.

\begin{figure*}[htbp]
\centering
\includegraphics[width=0.96\textwidth, height=0.5\textwidth]{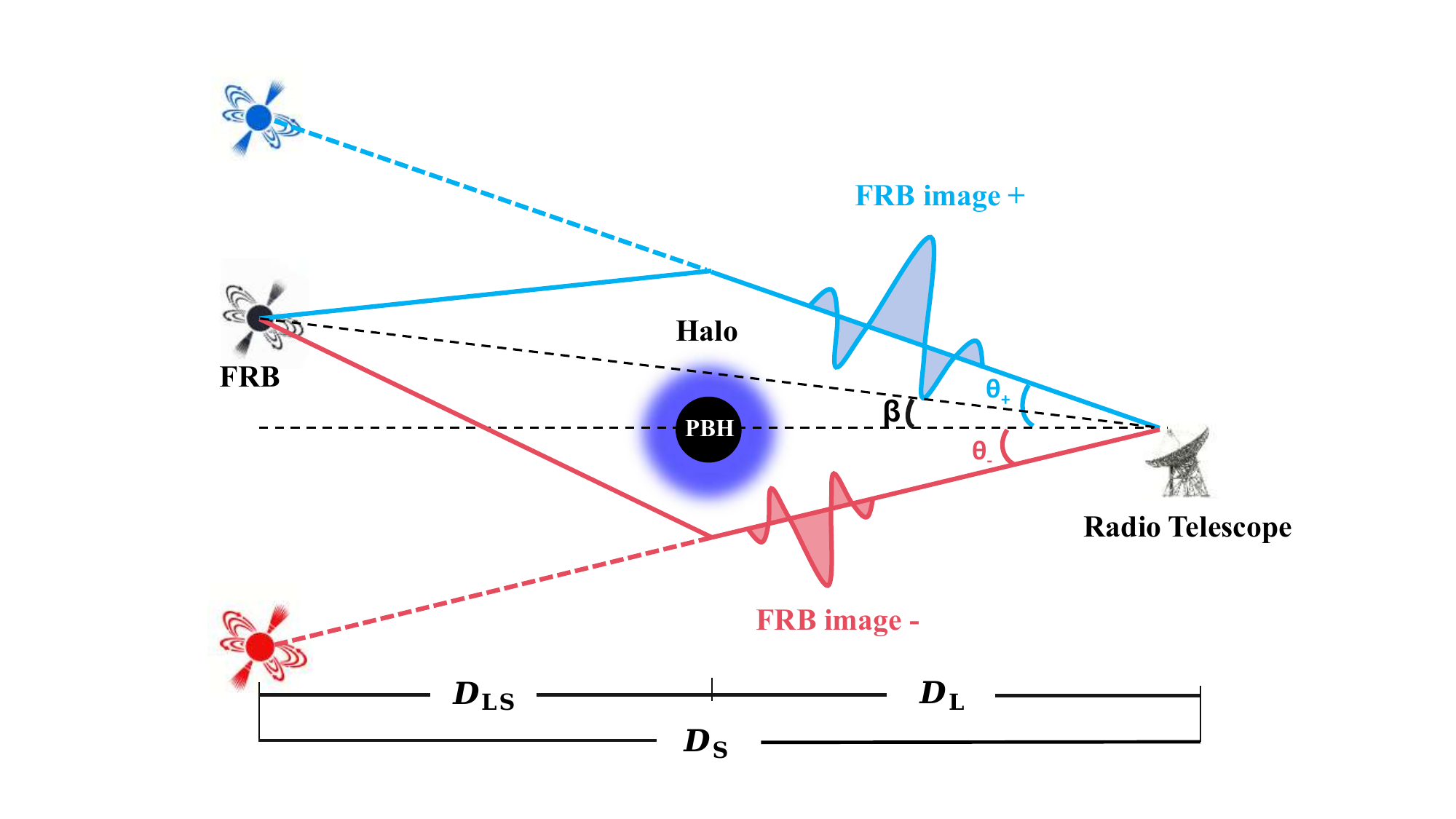}
\caption{Schematic diagram of the FRB being lensed considering dressed PBH.}
\label{fig1}
\end{figure*}

\section{Method}\label{sec:method}
In this section, we estimate the impact of the surrounding dark matter halo on PBH lensing, as schematically illustrated in Figure~\ref{fig1}. Based on optical depth theory, we derive the key analytical expression that converts any upper limit on the abundance of bare PBHs with monochromatic mass distribution into the corresponding constraint on dressed PBHs with arbitrary extended mass distributions.

\subsection{PBH dressed by Dark Matter Halo}\label{sub:A}
\begin{figure*}
    \centering
    \includegraphics[width=0.96\textwidth, height=0.7\textwidth]{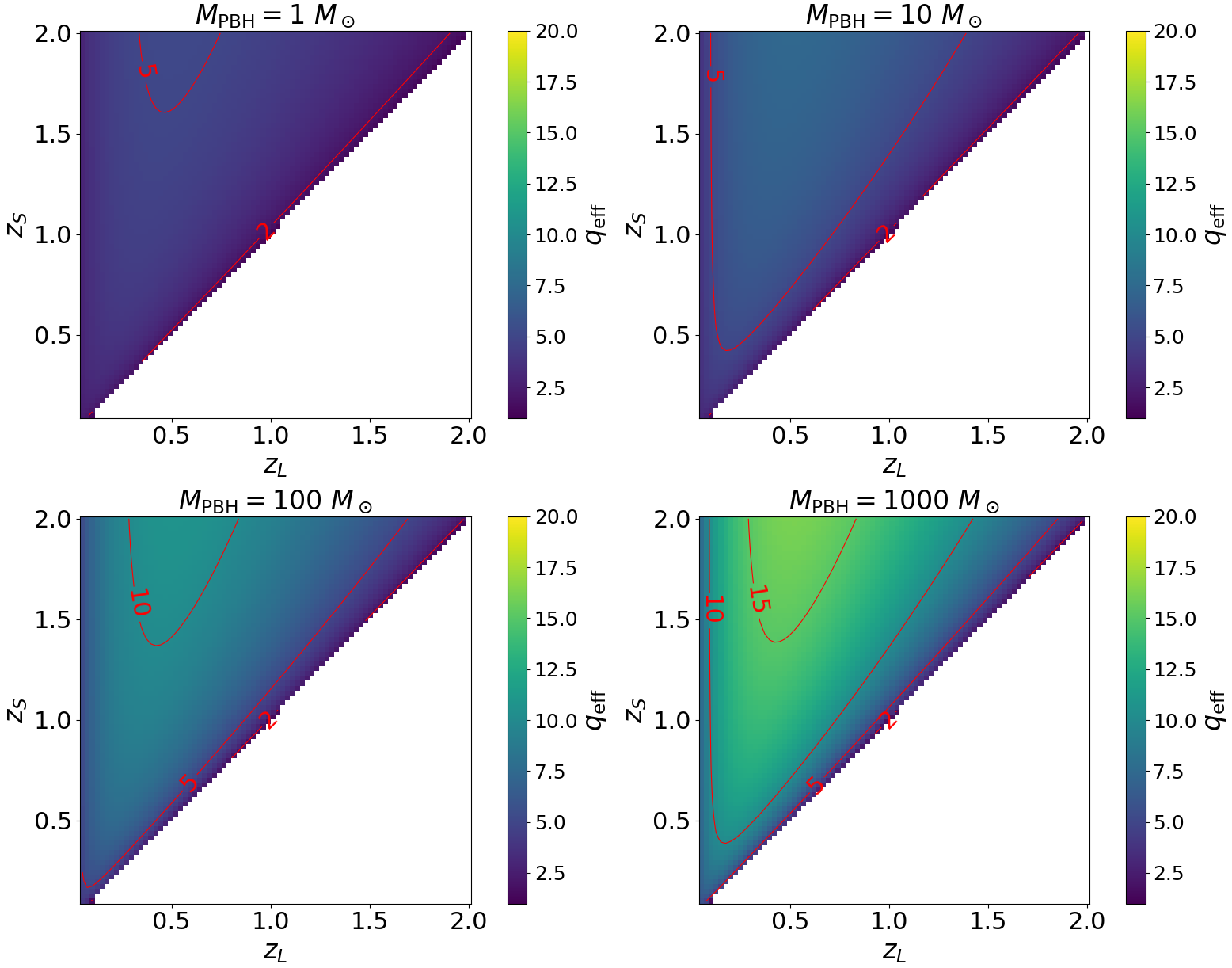}
    \caption{Mass ratio $q_{\rm eff}$ for PBHs with dark matter halos as a function of the bare PBH mass \(M_{\mathrm{PBH}}\in[1,\,10,\,100,\,1000]\,M_{\odot}\), with lens and source redshifts satisfying \(0 \le z_{\mathrm{L}} \le z_{\mathrm{S}} \le 2\). Contours are drawn for \(q_{\mathrm{eff}}\in[2,\,5,\,10,\,15]\).}
    \label{fig2}
\end{figure*}

In this paper, we consider the region of the density profile that formed when most PBHs were still isolated and accreting matter from the surrounding smooth dark matter background, thereby forming dark matter halos. This process occurs mainly during the matter-dominated era. The mass and radius of the dark matter halo grow with redshift up to about \(z_\mathrm{md} \approx 30\), with the following scaling relations~\cite{Mack:2006gz,Ricotti:2007au}
\begin{equation}
\begin{split}
&M_\mathrm{h}(z_\mathrm{md})\approx3 \left( \frac{1000}{1 + z_\mathrm{md}} \right) M_{\mathrm{PBH}},\\
&R_\mathrm{h}(z_\mathrm{md})\approx0.019 \, \mathrm{pc} \left( \frac{M_\mathrm{h}}{M_{\odot}} \right)^{1/3} \left( \frac{1000}{1 + z_\mathrm{md}} \right).
\label{eq2-1}
\end{split}
\end{equation}
According to \(N\)-body simulations, the halo profile can be well described by power-law distribution ~\cite{Boudaud:2021irr,Carr:2020mqm,Feng:2021qkj,Lavalle:2025rnx,Lavalle:2026fhx}
\begin{equation}
\rho_\mathrm{h}(r) = \rho_0 \left(R_\mathrm{h}/{r} \right)^{\alpha}~~~(\alpha=9/4),
\label{eq2-2}
\end{equation}
according to Eq.~(\ref{eq2-1}), $\rho_0$ can be obtained as
\begin{equation}
\rho_0=\frac{(3-\alpha)M_{\rm h}}{4\pi R_{\rm h}^3}\approx\frac{3-\alpha}{4\pi}\bigg(\frac{1+z_{\rm dm}}{19}\bigg)^3~M_{\odot}/{\rm pc}^3.
\label{eq2-3}
\end{equation}
Depending on assumptions about the underlying particle model, the inner density profile of the dark matter halo may deviate from the \(\propto r^{-9/4}\) behavior predicted by some analytic models~\footnote{We adopt the simplest theoretical model for the PBH halo structure. Other effects, such as those arising from different particle halo models, significant annihilation due to weak interactions, and dark matter decay, have been shown to have impact on the present theory~\cite{Adamek:2019gns,Boudaud:2021irr,Carr:2020mqm,Feng:2021qkj,Lavalle:2025rnx,Lavalle:2026fhx}. Nevertheless, these effects are not expected to significantly affect secondary phenomena such as gravitational lensing.}~\cite{Adamek:2019gns,Boudaud:2021irr,Carr:2020mqm,Feng:2021qkj,Lavalle:2025rnx,Lavalle:2026fhx}. To estimate the impact of the surrounding dark halo on PBH lensing, we consider the surface mass density \(\Sigma_{\rm h}\), defined as the line-of-sight projection of the lens's three-dimensional density, ignoring the truncation of the density profile beyond \(R_h\), as well as the truncation at the inner boundary
\begin{align}
\Sigma_\mathrm{h}(r) &= \int_{-\infty}^{\infty} dz \, \rho_0 \left( \frac{R_\mathrm{h}}{\sqrt{r^2 + z^2}} \right)^{\alpha} \nonumber\\
&= \rho_0 R_\mathrm{h} \sqrt{\pi} \frac{\Gamma((\alpha-1)/2)}{\Gamma(\alpha/2)} \left(R_\mathrm{h}/{r} \right)^{\alpha-1}.
\label{eq2-4}
\end{align}
From Eq.~(\ref{eq2-4}), the mean surface density of the dark matter halo within a radius \(r\) is obtained
\begin{align}
\bar{\Sigma}_\mathrm{h}(<r) &= \frac{1}{\pi r^2} \int_{0}^{r} \Sigma_\mathrm{h}(r') \, 2\pi r' \, dr' \nonumber\\
&= \frac{2\rho_0 R_\mathrm{h}\sqrt{\pi} }{3-\alpha}\frac{\Gamma((\alpha-1)/2)}{\Gamma(\alpha/2)}\left({R_\mathrm{h}}/{r} \right)^{\alpha-1}.
\label{eq2-5}
\end{align}
Similarly, the mean surface density of a PBH with mass \(M_{\mathrm{PBH}}\) is given by Dirac delta function \(\rho_{\rm PBH}(r) = M_{\rm PBH}\,\delta^3(r)\) as
\begin{equation}
\bar{\Sigma}_{\mathrm{PBH}}(<r) = \frac{M_{\mathrm{PBH}}}{\pi r^2}.
\label{eq2-6}
\end{equation}
Based on the above, we can compute the average convergence which delineates the regimes of strong and weak gravitational lensing~\cite{Narayan:1996ba}
\begin{equation}
\kappa = \frac{\bar{\Sigma}}{\Sigma_{\mathrm{cr}}},
\label{eq2-7}
\end{equation}
where \(\Sigma_{\mathrm{cr}}\) is the critical surface density as
\begin{equation}
\Sigma_{\mathrm{cr}} = \frac{1}{4\pi} \frac{D_\mathrm{S}}{D_\mathrm{L} D_\mathrm{LS}},
\label{eq2-8}
\end{equation}
where $D_{\rm S}$, $D_{\rm L}$ and $D_{\rm LS}$ represent the angular diameter distance to the source, to the lens, and between the source and the lens, respectively. From Eq.~(\ref{eq2-5}-\ref{eq2-6}), the average convergences of the halo and the PBH are then given by
\begin{equation}
\begin{split}
&\bar{\kappa}_\mathrm{h}(<r) = \frac{\bar{\Sigma}_\mathrm{h}(<r)}{\Sigma_{\mathrm{cr}}}, \\
&\bar{\kappa}_{\mathrm{PBH}}(<r) = \frac{M_{\mathrm{PBH}}}{\pi r^2 \Sigma_{\mathrm{cr}}}.
\end{split}
\label{eq2-9}
\end{equation}
The effect of the dark halo on PBH lensing can be estimated by comparing the average convergence within the PBH Einstein radius, and the total Einstein radius $r_{\rm E,tot}$ is then obtained by adding the halo contribution to the PBH mass as~\cite{Oguri:2022fir}
\begin{equation}
\bar{\kappa}_\mathrm{h}(< r_{\mathrm{E,tot}})+\bar{\kappa}_{\mathrm{PBH}}(< r_{\mathrm{E,tot}}) = 1.
\label{eq2-10}
\end{equation}
From Eq.~(\ref{eq2-10}), we define the total Einstein radius of the dressed PBH $r_{\rm E,tot}$ as that of a bare PBH with mass \(M_{\mathrm{PBH}}^{\mathrm{eff}}\) as 
\begin{equation}
r_{\rm E,PBH}(M_{\mathrm{PBH}}^{\mathrm{eff}})= r_{\rm E,tot}(M_{\mathrm{PBH}},z_{\mathrm{L}},\, z_{\mathrm{S}}).
\label{eq2-11}
\end{equation}
Based on the effective mass \(M_{\mathrm{PBH}}^{\mathrm{eff}}\), we define the mass ratio as
\begin{equation}
q_{\rm eff}\equiv\frac{M_{\mathrm{PBH}}^{\mathrm{eff}}}{M_{\mathrm{PBH}}}.
\label{eq2-12}
\end{equation}
Figure~\ref{fig2} shows the distribution of this ratio as a function of the bare PBH mass \(M_{\mathrm{PBH}}\in[1,\,10,\,100,\,1000]\,M_{\odot}\), with lens and source redshifts satisfying \(0 \leq z_{\mathrm{L}} \leq z_{\mathrm{S}} \leq 2\). In each panel, contours corresponding to \(q_{\mathrm{eff}}\in[2,\,5,\,10,\,15]\) are  displayed, confirming a monotonic increase of \(q_{\mathrm{eff}}\) with \(M_{\mathrm{PBH}}\). This trend arises from the halo mass-radius scaling relations: heavier PBHs accrete more dark matter, forming larger and more massive halos. Since the halo contribution grows faster with \(M_{\mathrm{PBH}}\) than does the PBH itself, the total Einstein radius increases more rapidly than its bare counterpart \(r_{\mathrm{E,PBH}}\propto\sqrt{M_{\mathrm{PBH}}}\). Consequently, \(M_{\mathrm{PBH}}^{\mathrm{eff}}\) grows super-linearly with respect to \(M_{\mathrm{PBH}}\), leading to the high \(q_{\mathrm{eff}}\) values observed. Furthermore, \(q_{\mathrm{eff}}\) exhibits a non-monotonic dependence on redshift. For very small source redshifts \(z_{\mathrm{S}}\), or when the lens redshift \(z_{\mathrm{L}}\) approaches \(z_{\mathrm{S}}\), the mass ratio tends to unity, indicating that the enhancement from the dark matter halo becomes negligible. As \(z_{\mathrm{S}}\) increases, the ratio attains a maximum for intermediate values of \(z_{\mathrm{L}}\). This trend is a direct consequence of the geometric factor entering the critical surface density \(\Sigma_{\mathrm{cr}} \propto D_{\mathrm{S}}/(D_{\mathrm{L}} D_{\mathrm{LS}})\): when the lens is located approximately midway between the source and the observer, \(\Sigma_{\mathrm{cr}}\) reaches its minimum, maximizing the halo contribution; conversely, the halo enhancement is suppressed when the lens is too close to either the observer or the source.

\subsection{\texorpdfstring{Constraints on $f_{\mathrm{PBH}}$ from Lensing Effect}{Constraints on f PBH from the Lensing Effect}}\label{sub:B}
As shown in Figure~\ref{fig1}, in the absence of a dark matter halo, the bare PBH is treated as a point mass, and its Einstein radius is given by
\begin{equation}
\begin{split}
r_{\rm E,PBH}(M_{\mathrm{PBH},},z_{\mathrm{L}},\, z_{\mathrm{S}})=D_{\rm L}\theta_\mathrm{E}\\
= \sqrt{\frac{4M_{\mathrm{PBH}} D_\mathrm{L}D_\mathrm{LS}}{D_\mathrm{S}}}.
\end{split}
\label{eq3-1}
\end{equation}
For a source lensed by a PBH, two images are produced, and the time delay between their arrivals is given by
\begin{equation}
\begin{split}
&\Delta t(M_{\mathrm{PBH}},\lambda) = 4 M_{\mathrm{PBH}} (1 + z_\mathrm{L})\times \\
&\bigg[\frac{y}{2} \sqrt{y^2 + 4} +\ln\left( \frac{\sqrt{y^2 + 4} + y}{\sqrt{y^2 + 4} - y} \right) \bigg],
\end{split}
\label{eq3-2}
\end{equation}
where \(\lambda\equiv[z_{\mathrm{L}},\, z_{\mathrm{S}},\, y]\) denotes the set of parameters characterizing the PBH and source, \(y \equiv \beta/\theta_\mathrm{E}\) is a dimensionless parameter characterizing the source position, and \(\beta\) denotes the angular position of the source. The lensing cross section due to a bare PBH lens is given by an annulus between the maximum and minimum source position
\begin{equation}
\begin{aligned}
\sigma_{\rm L}(M_{\mathrm{PBH}},\lambda, \Lambda) = \pi r_{\mathrm{E,PBH}}^2\bigl[y_{\mathrm{max}}^2(\Lambda_1) - y_{\mathrm{min}}^2(\Lambda_2)\bigr] \\
= \frac{4\pi M_{\mathrm{PBH}} D_{\mathrm{L}} D_{\mathrm{LS}}}{D_{\mathrm{S}}} \bigl[y_{\mathrm{max}}^2(\Lambda_1) - y_{\mathrm{min}}^2(\Lambda_2) \bigr],
\end{aligned}
\label{eq3-3}
\end{equation}
where \(\Lambda \equiv [\Lambda_1, \Lambda_2]\) denotes the selection-effect parameters that determine whether a lensing system is detectable. These parameters depend on both the instrument properties and the intrinsic features of the source signal, e,g., the minimum impact parameter $y_{\min}(\Lambda_2)$ of microlensed FRB is the function of its signal width $\omega$. For a single source, the optical depth contributed by a bare PBH can be written as
\begin{equation}
\begin{aligned}
\tau^{\rm wo}_{\rm MMD}(f_{\text{PBH}},M_{\text{PBH}},z_\mathrm{S}) = \int_{0}^{z_\mathrm{S}} d\chi(z_\mathrm{L})(1 + z_\mathrm{L})^2 \times\\
n_\mathrm{PBH}(f_{\text{PBH}},M_{\text{PBH}})\sigma_{\rm L}(M_{\text{PBH}}, \lambda, \Lambda) 
\end{aligned}
\label{eq3-4}
\end{equation}
where $n_\mathrm{PBH}(f_{\text{PBH}}, M_{\mathrm{PBH}})$ is the comoving number
density of the PBHs with MMD
\begin{equation}
n_\mathrm{PBH}(f_{\text{PBH}},M_{\text{PBH}}) = \frac{f_{\text{PBH}} \, \Omega_{\text{DM}} \rho_c}{M_{\mathrm{PBH}}}
\label{eq3-5}
\end{equation}
where $\rho_c$ is the critical density of the universe, and $\Omega_{\rm DM}$ is the dark matter density parameter.
The optical depth for a dressed PBH, i.e., a PBH surrounded by dark matter halo, can be computed as
\begin{equation}
\begin{aligned}
\tau_{\text{MMD}}^{\mathrm{w}}(f_{\text{PBH}}, M_{\text{PBH}}, z_\mathrm{S}) = \int_0^{z_\mathrm{S}} d\chi(z_\mathrm{L}) \, (1 + z_\mathrm{L})^2 \times\\ n_{\text{PBH}}(M_{\mathrm{PBH}},f_{\text{PBH}}) \, \sigma_{\rm L}(M_{\text{PBH}}^{\text{eff}}, \lambda,\Lambda).
\end{aligned}
\label{eq3-6}
\end{equation}
It should be emphasized that the halo enhances the lensing cross section but leaves the PBH comoving number density unaffected. Secondary effects, e.g., those from binary black hole mergers~\cite{Sasaki:2016jop,Ali-Haimoud:2017rtz,Chen:2018czv}, could potentially alter the density, but their magnitude is negligibly small. Above formalism is only valid for MMD
\begin{equation}
\psi(M_{\mathrm{PBH}},m) = \delta(m-M_{\text{PBH}}),
\label{eq3-7}
\end{equation}
where $\delta(m-M_{\mathrm{PBH}})$ represents the $\delta$-function at $M_{\rm PBH}$. However, MMD is an oversimplified model that lacks realistic significance, as PBHs are not expected to form at a single epoch with a unique mass~\cite{Green:2004wb,Niemeyer:1997mt,Carr:2017jsz,Bellomo:2017zsr}. Since different inflation models yield distinct primordial curvature power spectra and hence different EMDs, it is essential to derive constraints on PBH with physically motivated EMD that reflect realistic formation scenario. For the EMD, the optical depth for a lensing event can be expressed as
\begin{equation}
\begin{split}
\tau_{\mathrm{EMD}}^{\mathrm{wo}}(f_{\text{PBH}}, \boldsymbol p_{\rm mf},z_\mathrm{S}) = \int dm \int_0^{z_\mathrm{S}} d\chi(z_\mathrm{L}) (1 + z_\mathrm{L})^2\times\\
\frac{dn_\mathrm{PBH}(f_{\text{PBH}}, \boldsymbol p_{\rm mf},m)}{dm}\sigma_{\rm L}(m, \lambda, \Lambda),
\end{split}
\label{eq3-8}
\end{equation}
where $\frac{dn_{\mathrm{PBH}}(f_{\mathrm{PBH}}, \boldsymbol p_{\rm mf}, m)}{dm}$ is the comoving number density of the PBHs at EMD $\psi(\boldsymbol p_{\rm mf},m)$
\begin{equation}
\frac{dn_{\mathrm{PBH}}(f_{\mathrm{PBH}}, \boldsymbol p_{\rm mf}, m)}{dm} =
\frac{f_{\mathrm{PBH}}\Omega_{\text{DM}} \rho_c}{m}\psi(\boldsymbol p_{\rm mf},m).
\label{eq3-9}
\end{equation}
Accordingly, the optical depth for a PBH dressed with a dark matter halo, under an EMD, can be written as
\begin{equation}
\begin{split}
\tau_{\mathrm{EMD}}^{\mathrm{w}}(f_{\text{PBH}}, \boldsymbol p_{\rm mf},z_\mathrm{S}) = \int dm \int_0^{z_\mathrm{S}} d\chi(z_\mathrm{PBH})(1 + z_\mathrm{L})^2\times \\
\frac{dn_\mathrm{PBH}(f_{\text{PBH}}, \boldsymbol p_{\rm mf},m)}{dm}\sigma_{\rm L}(m^{\rm eff}, \lambda, \Lambda),
\end{split}
\label{eq3-10}
\end{equation}
where \(m^{\rm eff}\) is the effective mass corresponding to the bare mass \(m\).

According to Poisson law, the probability for null detecting lensed events among a total of \(N_{\mathrm{obs}}\) independent observational events $d\in[d_1,d_2,.....d_{N_{\rm obs}}]$ is given by
\begin{equation}
P_i(f_{\text{PBH}}) = \exp(-\tau_i(f_{\text{PBH}}))
\label{eq3-11}
\end{equation}
where \(\tau_i\) denotes the optical depth, defined independently of the specific lens model. 
For a null search of lensed signals, the upper limit on \(f_{\mathrm{PBH}}\) is obtained by requiring that the non-detection be consistent with the hypothesis that PBHs constitute a fraction \(f_{\mathrm{PBH}}\) of the dark matter at the \(100\Pi\%\) confidence level
\begin{equation}
\begin{split}
&P_{\text{obs}}(f_{\text{PBH}}) = \prod_{i}^{N_{\rm obs}}P_i(f_{\text{PBH}})=\\
&\exp\left(-\sum_{i=1}^{N_{\text{obs}}} \tau_i(f_{\text{PBH}})\right)\geq 1 - \Pi.
\end{split}
\label{eq3-12}
\end{equation}
To distinguish the constraints obtained from the four different optical depth cases $\tau\equiv[\tau_{\rm MMD}^{\rm wo}, \tau_{\rm MMD}^{\rm w}, \tau_{\rm EMD}^{\rm wo}, \tau_{\rm EMD}^{\rm w}]$, we label the corresponding upper limits of \(f_{\mathrm{PBH}}\) from Eq.~(\ref{eq3-12}) as
\begin{equation}
\begin{split}
f_{\rm PBH,MMD}\equiv[f_{\rm PBH,MMD}^{\rm wo}, f_{\rm PBH,MMD}^{\rm w}],\\
f_{\rm PBH,EMD}\equiv[f_{\rm PBH,EMD}^{\rm wo}, f_{\rm PBH,EMD}^{\rm w}].
\end{split}
\label{eq3-13}
\end{equation}
For a single lensing system, the optical depths in the halo-dressed case under two types of mass distributions are
\begin{equation}
\left\{
\begin{aligned}
& \tau_{\mathrm{MMD}}^{\mathrm{w}}\left(f_{\mathrm{PBH,MMD}}^{\mathrm{w}}, M_{\mathrm{PBH}}\right) \\
& \quad = f_{\mathrm{PBH,MMD}}^{\mathrm{w}} \, \tau_{\mathrm{MMD}}^{\mathrm{w}}\left(f_{\mathrm{PBH,MMD}}^{\mathrm{w}} = 1, M_{\mathrm{PBH}}\right), \\[10pt]
& \tau_{\mathrm{EMD}}^{\mathrm{w}}\left(f_{\mathrm{PBH,EMD}}^{\mathrm{w}},\boldsymbol p_{\rm mf}\right) \\
& \quad = f_{\mathrm{PBH,EMD}}^{\mathrm{w}} \, \tau_{\mathrm{EMD}}^{\mathrm{w}}\left(f_{\mathrm{PBH,EMD}}^{\mathrm{w}} = 1, \boldsymbol p_{\rm mf}\right).
\end{aligned}
\right.
\label{eq3-14}
\end{equation}
From the relationship between the optical depths for the MMD and EMD with halo in Eq.~(\ref{eq3-14}), we obtain
\begin{equation}
\begin{split}
\tau_{\mathrm{EMD}}^{\mathrm{w}} (f_{\mathrm{PBH,EMD}}^{\mathrm{w}} = 1, \boldsymbol p_{\rm mf}) = \int_{0}^{+\infty} dm\; \psi(\boldsymbol p_{\rm mf},m) \\
\times \;\tau_{\mathrm{MMD}}^{\mathrm{w}} (f_{\mathrm{PBH,MMD}}^{\mathrm{w}} = 1, m).
\end{split}
\label{eq3-15}
\end{equation}
In addition, we can respectively obtain the conservative upper limits of $f_{\mathrm{PBH}}$ from Eq.~(\ref{eq3-12}) for the MMD and EMD scenarios, respectively.
\begin{widetext}
\begin{equation}
\begin{cases}
f_{\mathrm{PBH,MMD}}^{\mathrm{w}}(M_{\mathrm{PBH}}) = 
   \dfrac{-\ln(1-\Pi)}
        {\displaystyle\sum_{i=1}^{N_{\mathrm{obs}}} 
         \tau_{\mathrm{MMD}}^{\mathrm{w}}(f_{\mathrm{PBH,MMD}}^{\mathrm{w}}=1,M_{\mathrm{PBH}})}, 
   \\
f_{\mathrm{PBH,EMD}}^{\mathrm{w}}(\boldsymbol p_{\rm mf}) = 
   \dfrac{-\ln(1-\Pi)}
        {\displaystyle\sum_{i=1}^{N_{\mathrm{obs}}} 
         \tau_{\mathrm{EMD}}^{\mathrm{w}}(f_{\mathrm{PBH,EMD}}^{\mathrm{w}}=1, \boldsymbol p_{\rm mf})} .
         \label{eq3-16}
\end{cases}
\end{equation}
\end{widetext}
From the above Eq.~(\ref{eq3-15}-\ref{eq3-16}), we can obtain
\begin{widetext}
\begin{equation}
\frac{f_{\mathrm{PBH,EMD}}^{\mathrm{w}}(\boldsymbol p_{\rm mf})}{f_{\mathrm{PBH,MMD}}^{\mathrm{w}}(M_{\rm PBH})}
= \frac{\displaystyle\sum_{i=1}^{N_{\mathrm{obs}}}\tau_{\mathrm{MMD}}^{\mathrm{w}}(f_{\mathrm{PBH,MMD}}^{\mathrm{w}}=1, M_{\mathrm{PBH}})}
{\displaystyle\sum_{i=1}^{N_{\mathrm{obs}}}\tau_{\mathrm{EMD}}^{\mathrm{w}}(f_{\mathrm{PBH,EMD}}^{\mathrm{w}}= 1, \boldsymbol p_{\rm mf})}
= \frac{\displaystyle\sum_{i=1}^{N_{\mathrm{obs}}} \tau_{\mathrm{MMD}}^{\mathrm{w}}(f_{\mathrm{PBH,MMD}}^{\mathrm{w}}=1, M_{\mathrm{PBH}})}
{\displaystyle\sum_{i=1}^{N_{\mathrm{obs}}}\int_0^\infty dm ~\tau_{\mathrm{MMD}}^{\mathrm{w}}(f_{\mathrm{PBH,MMD}}^{\mathrm{w}}=1, m)\psi(\boldsymbol p_{\rm mf}, m)}.
\label{eq3-17}
\end{equation}
\end{widetext}
Finally, we can integrate Eq.~(\ref{eq3-17}) with the same mass distribution $\psi(\boldsymbol p_{\rm mf}, M_{\rm PBH})$ over $M_{\rm PBH}$ to obtain
\begin{widetext}
\begin{equation}
\int_{0}^{\infty} dM_{\rm PBH} \, \frac{f_{\mathrm{PBH,EMD}}^{\mathrm{w}}(\boldsymbol p_{\rm mf}) \; \psi(\boldsymbol p_{\rm mf}, M_{\rm PBH})}{f_{\mathrm{MMD}}^{\mathrm{w}}(M_{\mathrm{PBH}})}= 
\frac{\displaystyle\int_{0}^{\infty} dM_{\rm PBH} \; \sum_{i=1}^{N_{\mathrm{obs}}} \tau_{\mathrm{MMD}}^{\mathrm{w}}(f_{\mathrm{PBH,MMD}}^{\mathrm{w}}=1, M_{\rm PBH}) \; \psi(\boldsymbol p_{\rm mf}, M_{\rm PBH})}{\displaystyle\sum_{i=1}^{N_{\mathrm{obs}}} \int_{0}^{\infty} dm \; \tau_{\mathrm{MMD}}^{\mathrm{w}}(f_{\mathrm{PBH,MMD}}^{\mathrm{w}}=1, m) \; \psi(\boldsymbol p_{\rm mf},m)} = 1.
\label{eq3-18}
\end{equation}
\end{widetext}
Rearranging Eq.~(\ref{eq3-18}) yields the following simplified expression as
\begin{equation}
f_{\mathrm{PBH,EMD}}^{\mathrm{w}}(\boldsymbol p_{\rm mf}) = \frac{1}{\displaystyle\int_{0}^{+\infty}\frac{\psi(\boldsymbol p_{\rm mf}, M_{\rm PBH})}{f_{\mathrm{PBH,MMD}}^{\mathrm{w}}(M_{\mathrm{PBH}})} \, dM_{\rm PBH}}.
\label{eq3-19}
\end{equation}
The above transformation relation is independent of the specific formation history of PBH dark matter halos, and is applicable not only to FRBs but also to any microlensing scenario. Furthermore, even if baryonic accretion alters the mass distribution of PBHs--as discussed in some works~\cite{Yuan:2023bvh,Dayal:2025aiv}--this does not affect our conclusion, since the net effect of accretion is merely to recast the mass distribution into an effective one. This constitutes a universal result, analogous to those obtained by~\citet{Carr:2017jsz} and~\citet{Zhou:2023awq} in the bare PBH case
\begin{equation}
f_{\mathrm{PBH,EMD}}^{\mathrm{wo}}(\boldsymbol p_{\rm mf}) = \frac{1}{\displaystyle\int_{0}^{+\infty}\frac{\psi(\boldsymbol p_{\rm mf}, M_{\rm PBH})}{f_{\mathrm{PBH,MMD}}^{\mathrm{wo}}(M_{\mathrm{PBH}})} \, dM_{\rm PBH}}.
\label{eq3-20}
\end{equation}

\section{Forecasts for PBH constraints}\label{sub:three}
For the purpose of probing PBHs through lensing effects, we forecast the constraints on PBH abundance achievable with future FRB observations based on above reduced analytical formulation. In this section, we separately demonstrate the simulation of FRBs and the application of simulated data to constrain PBH properties across a range of scenarios.

\subsection{Mock FRB Samples}\label{sub1}
Although current constraints on PBHs are weak owing to the paucity of FRBs, the rapid expansion of FRB detections is now changing the landscape. Wide-field radio telescopes such as CHIME, FAST, ASKAP, and DSA are producing ever-growing samples, e.g., the CHIME/FRB Catalog 2, which already contains 4539 bursts from 3641 distinct sources~\cite{TheCHIMEFRB:2026nji}, has been used to place upper limits on the PBH abundance at the level of $\sim10^{-1}$~\cite{Zhou:2026flq}. For future forecasts, CHIME has a system equivalent flux density (SEFD) of \(80\)--\(90\ \text{Jy}\), a minimum detectable flux \(S_{\text{min}} \approx 0.1\ \text{Jy}\), and a field of view (FoV) of about \(140\ \text{deg}^2\)~\cite{CHIMEFRB:2018mlh}. It is expected to observe approximately \(10^5\) bursts over twenty year operational period, i.e., about \(5 \times 10^3\) per year~\cite{Connor:2022bwl,He:2026nfs}. In addition, the Bustling Universe Radio Survey Telescope in Taiwan (BURSTT) has a field of view of \(5000\) square degrees and an SEFD of \(600\ \text{Jy}\) (corresponding to \(S_{\text{min}} \approx 0.94\ \text{Jy}\))~\cite{Lin:2022wgp}, and is expected to detect about \(5 \times 10^5\) bursts in ten years~\cite{He:2026nfs}. The Square Kilometre Array Phase 2 (SKA2), with its superior SEFD (\(0.2\)--\(0.3\ \text{Jy}\) in the low band, \(0.1\)--\(0.2\ \text{Jy}\) in the mid band, corresponding to \(S_{\text{min}}\) as low as \(10^{-4}\ \text{Jy}\)), despite a smaller field of view (about \(100\)--\(150\) square degrees)~\cite{Braun:2019gdo}, is expected to detect as many as \(10^6\)--\(10^7\) bursts over ten years~\cite{He:2026nfs}. These data show that even a single facility can conservatively reach \(10^5\) FRB detections within a decade; combining multiple observatories (e.g., the FAST array expansion~\cite{peng2024}, DSA-2000~\cite{Hallinan:2019qyo}, The Canadian Hydrogen Observatory and Radio-transient Detector (CHORD)~\cite{Vanderlinde:2019tjt}) will further boost the detection rate. Hence, a sample of \(10^5\) FRBs in ten years is a realistic expectation. Based on such a large sample, we derive upper limits on the PBH abundance under different scenarios. In the following analysis, we adopt a constant-density redshift distribution (CRD) for FRBs, denoted by \(P_{\rm CRD}(z)\), to forecast the PBH constraints. The CRD is expressed as~\citep{Munoz:2016tmg, Laha:2018zav,Oguri:2022fir,Zhou:2021tvp}
\begin{equation}
P_{\rm CRD}(z)=\frac{1}{Z_{\rm c}}\frac{\chi^2(z)e^{-d^2_{\rm L}(z)/[2d^2_{\rm L}(z_{\rm cut})]}}{(1+z)H(z)},
\label{eq4-1}
\end{equation}
where $Z_{\rm c}$ is a normalization factor to ensure that $P_{\rm CRD}(z)$ integrates to unity, $d_{\rm L}(z)$is the luminosity distance, and $z_{\rm cut}=0.5$ is Gaussian cutoff in the FRB redshift distribution due to an instrumental signal-to-noise threshold.

For microlensing of FRBs, the critical value $\Lambda_1=[R_{\rm f, max}, T_{\rm obs}]$ and the width $\Lambda_2=w$ of the observed signal determine the maximum and minimum value of impact parameter in optical depth. To ensure that both microlensed signals are detectable, the first maximum impact parameter \(y_{\max,1}(\Lambda_1=R_{\rm f,max})\) is determined by requiring that the flux ratio between two signals be smaller than threshold \(R_{\mathrm{f,max}}\)
\begin{equation}
y_{\max,1}(\Lambda_1=R_{\rm f,max})=R_{\rm f,max}^{1/4}-R_{\rm f,max}^{-1/4},
\label{eq4-2}
\end{equation}
we adopt \(R_{\mathrm{f,max}}=5\) for the simulated FRB sample, following previous works~\cite{Munoz:2016tmg, Laha:2018zav,Oguri:2022fir, Liao:2020wae,Zhou:2021tvp}. In addition, the second maximum impact parameter \(y_{\max,2}(\Lambda_1=T_{\rm obs})\) and the minimum impact parameter \(y_{\min}(\Lambda_2=w)\) are obtained by imposing the condition that the time delay between the two microlensed signals be smaller than the observation duration \(T_{\rm obs}\) to ensure that both images arrive within the observing window, and also larger than the signal width \(w\) to avoid temporal overlap
\begin{equation}
w\leq\Delta t(M_{\mathrm{PBH}},\lambda)\leq T_{\rm obs}.
\label{eq4-3}
\end{equation}
We considered two average widths $\bar{w}\in[0.1, 1]~{\rm ms}$ and $T_{\rm obs}=1~\rm min$ in our subsequent analysis. The final maximum impact parameter is taken as 
\begin{equation}
y_{\max} = \min(y_{\max,1}, y_{\max,2}).
\label{eq4-4}
\end{equation}

\begin{figure}
\includegraphics[width=0.45\textwidth, height=0.36\textwidth]{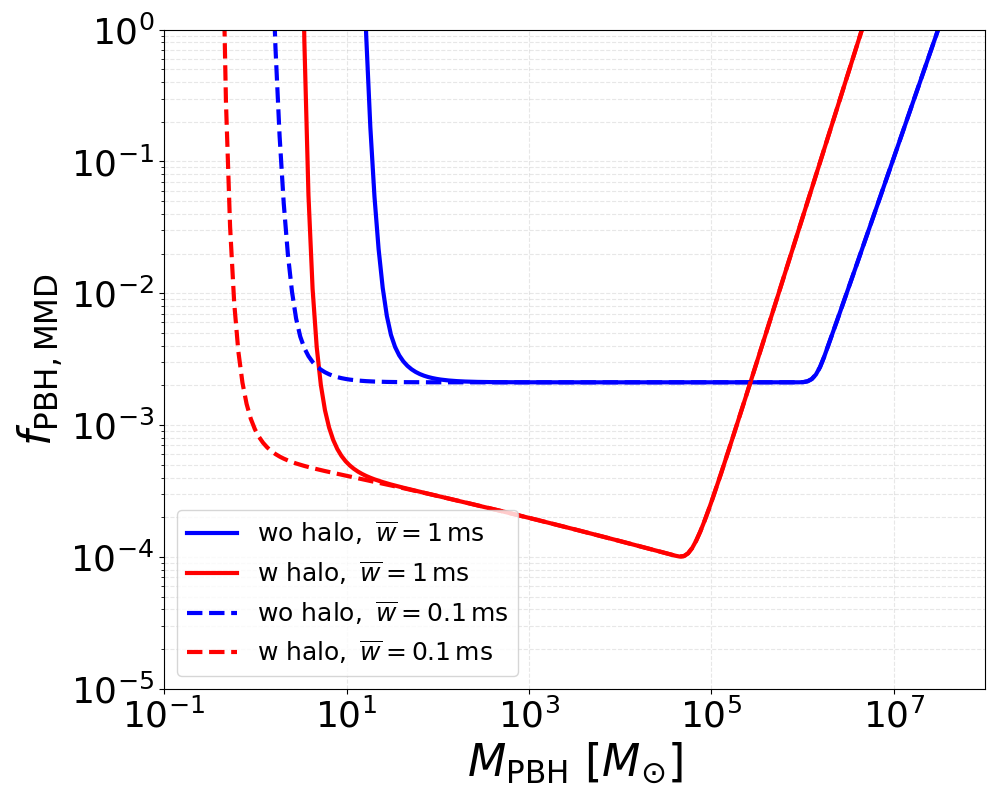}
\caption{\label{fig:3}Upper limits at 95\% confidence level on the PBH abundance \(f_{\mathrm{PBH,MMD}}\) for bare and halo-dressed PBHs under MMD, inferred from \(10^5\) mock FRB events.}
\end{figure}

\subsection{Results}\label{sub2}
In Figure~\ref{fig:3}, we demonstrate the constraints on the PBH dark matter fraction for both bare PBHs and the MMD. The solid lines and dashed lines represent the constraints on $f_{\rm PBH, MMD}$ from $10^5$ FRBs with $\bar{w}=1~\rm ms$ and $\bar{w}=0.1~\rm ms$, respectively. As expected, a smaller \(\bar{w}\) enables constraints on lower PBH masses in the \(M_{\mathrm{PBH}}\)-\(f_{\mathrm{PBH}}\) plane, while the high-mass cutoff is the same for both cases because the observation duration \(T_{\rm obs}\) is fixed to a common value. For both \(\bar{w}=1\ \mathrm{ms}\) and \(0.1\ \mathrm{ms}\), the strongest upper limits at $95\%$ confidence level on \(f_{\mathrm{PBH}}\) are \(2.1\times10^{-3}\) for bare PBHs and \(1.0\times10^{-4}\) for halo-dressed PBHs, indicating that the presence of a dark matter halo strengthens the constraints by approximately a factor of 20. Moreover, the nonlinear enhancement of constraints by halo-dressed PBHs displayed in Figure~\ref{fig:3} is a direct consequence of the superlinear \(M_{\mathrm{PBH}}^{\mathrm{eff}}\)-\(M_{\mathrm{PBH}}\) relation discussed earlier. 

\begin{figure*}
\centering
    \includegraphics[width=0.48\textwidth, height=0.38\textwidth]{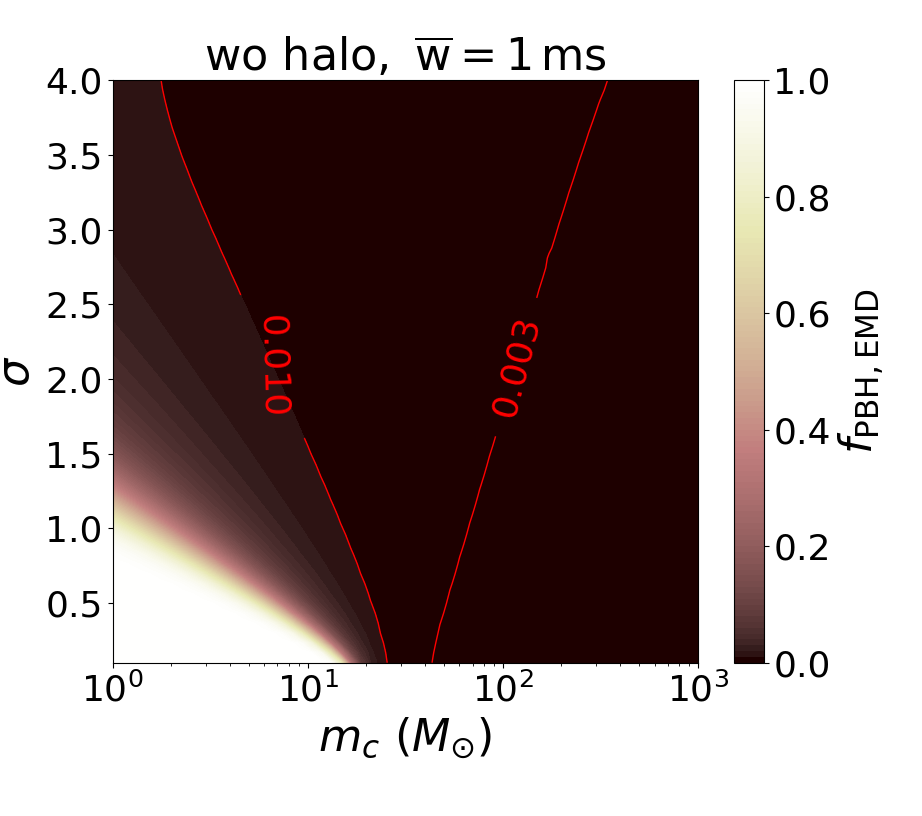}
    \includegraphics[width=0.48\textwidth, height=0.38\textwidth]{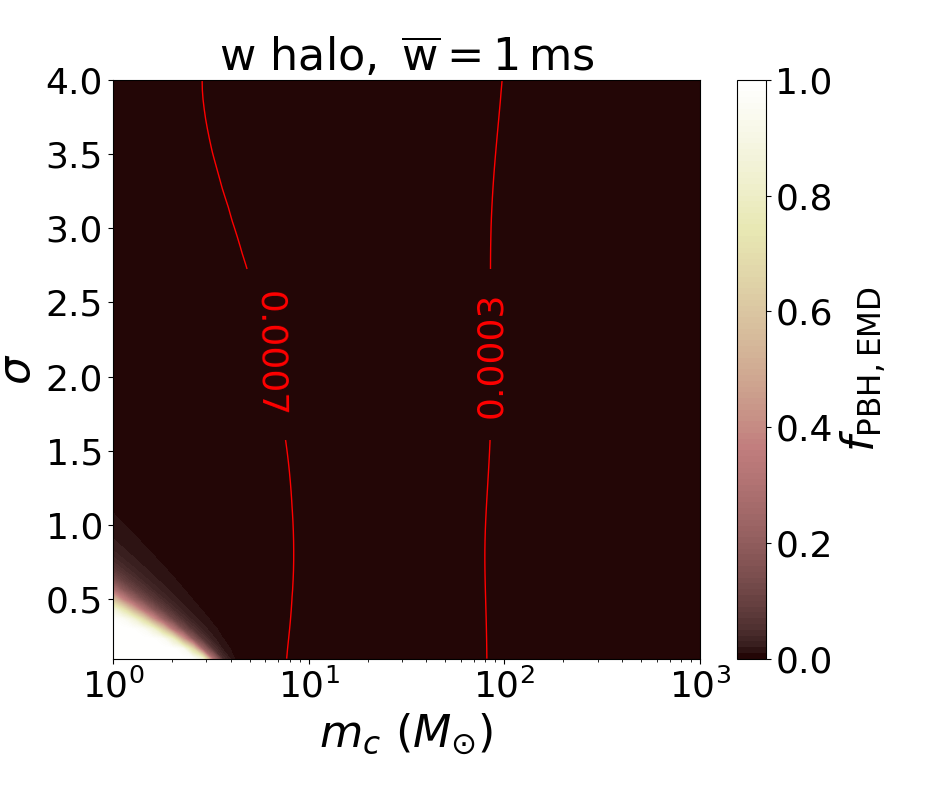}
    \includegraphics[width=0.48\textwidth, height=0.38\textwidth]{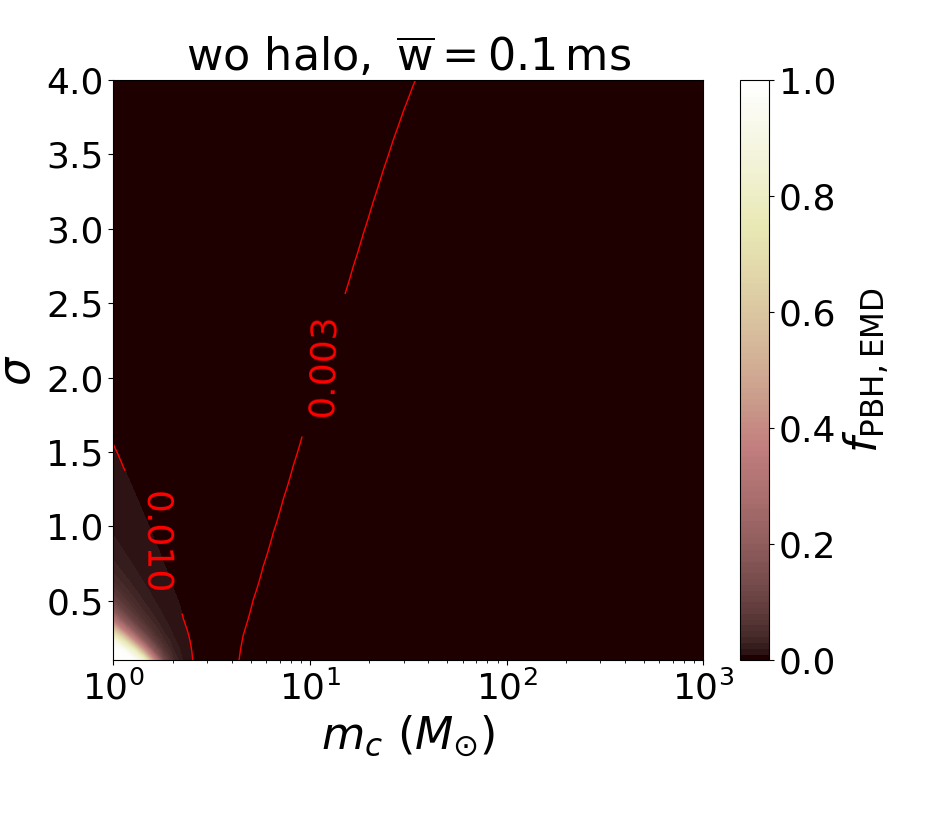}
    \includegraphics[width=0.48\textwidth, height=0.38\textwidth]{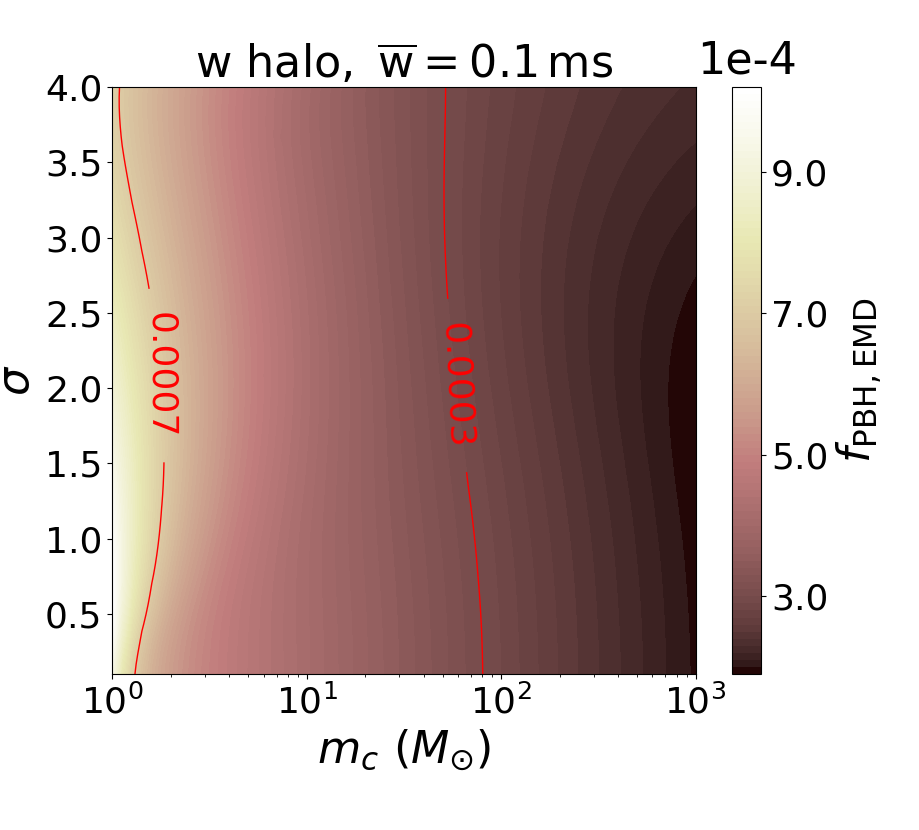}
    \caption{$95\%$ confidence level upper limits on \(f_{\mathrm{PBH,EMD}}\) from \(10^5\) FRB lensing null searches, assuming a log-normal mass function (\(m_{\rm c}\in[1,10^3]\,M_\odot\), \(\sigma\in[0.1,4.0]\)). 
    \textbf{Upper left:} Results for bare PBHs with \(\bar{w}=1\,\mathrm{ms}\), and contours are plotted at \(f_{\mathrm{PBH}}=10^{-2}\) and \(3\times10^{-3}\).\textbf{Upper right:} Results for halo-dressed PBHs with \(\bar{w}=1\,\mathrm{ms}\), and contours at \(f_{\mathrm{PBH}}=7\times10^{-3}\) and \(3\times10^{-3}\);
    \textbf{Lower left:} Same as upper left, but with \(\bar{w}=0.1\,\mathrm{ms}\);
    \textbf{Lower right:} Same as upper right, but with \(\bar{w}=0.1\,\mathrm{ms}\).}
    \label{fig:4}
\end{figure*}

Next, we demonstrate, within the framework of an EMD, the effect of dark matter halos on the PBH abundance constraints derived from unlensed FRB observations. Specifically, we adopt a log-normal mass function to characterize the EMD~\cite{Carr:2017jsz, Bellomo:2017zsr}:
\begin{equation}
\psi(\boldsymbol p_{\rm mf},m)=\frac{1}{\sqrt{2\pi}\sigma m}\exp\bigg[-\frac{\ln^2(m/m_{\rm c})}{2\sigma^2}\bigg],
\label{eq4-5}
\end{equation}
where $\boldsymbol p_{\rm mf}=[\sigma,m_{\rm c}]$, $m_{\rm c}$ and $\sigma$ denote the peak mass of $m\psi(\boldsymbol p_{\rm mf},m)$ and the width of mass spectrum, respectively. This mass function provides a good approximation when bare PBHs are produced from a smooth symmetric peak in the inflationary power spectrum, and has been demonstrated numerically and analytically under the slow-roll approximation. It thus serves as a representative of a broad class of extended mass functions, and is also widely used in the population analysis of binary black hole mergers from gravitational wave observations~\cite{Hutsi:2020sol,Franciolini:2021tla,Chen:2022fda,Bouhaddouti:2026jgc}. Figure~\ref{fig:4} shows the results for this mass distribution, with the peak mass parameter \(m_c\) ranging from \(1\) to \(10^3\,M_\odot\) and the width parameter \(\sigma\) from \(0.1\) to \(4\). For the left panel (bare PBHs), we use Eq.~(\ref{eq3-20}) to convert the MMD upper limits to the EMD case; for the right panel (dressed PBHs), we employs our key transformation given by Eq.~(\ref{eq3-19}) to perform the same conversion. We find that, for the log-normal mass function, a larger characteristic mass \(m_c\) concentrates most of the mass at high masses, resulting in a larger lensing optical depth and hence stronger constraints on \(f_{\mathrm{PBH,EMD}}\). Whereas increasing \(\sigma\) at large \(m_c\) introduces more low-mass PBHs, which reduce the optical depth and thus weaken the constraints. The strongest $95\%$ confidence level upper limits on \(f_{\mathrm{PBH,EMD}}\) for bare and halo-dressed PBHs are \(2.1\times10^{-3}\) and \(1.9\times10^{-4}\), respectively, at \(\bar{w}=1\,\mathrm{ms}\). At \(\bar{w}=0.1\,\mathrm{ms}\), the corresponding limits are \(2.1\times10^{-3}\) and \(1.9\times10^{-4}\). In agreement with the results presented in Figs.~\ref{fig:3}, and similar to constraints from other independent probes~\cite{Urrutia:2023mtk,Cai:2022kbp}, the presence of dark matter halos significantly boosts the gravitational lensing cross section of PBHs, leading to constraints on \(f_{\mathrm{PBH}}\) that are roughly an order of magnitude stronger for halo-dressed PBHs than for bare ones.

\section{Conclusion and Disscussions}\label{sub:four}

PBHs have drawn broad interest across cosmology and astrophysics: they are one of dark matter candidates, potential sources of gravitational waves from binary mergers observed by LIGO-Virgo-KAGRA, and possible seeds for the SMBHs detected by JWST. FRBs, bright millisecond-duration radio transients of unknown origin, have rapidly become a powerful probe in astronomy, with their microlensing effect offering a clean and sensitive method to constrain PBHs, particularly in the mass range above the stellar-mass window. In this paper, we derive a transformation that converts upper limits on the abundance of bare PBHs with MMD into constraints on dressed PBHs with EMD. Under this framework, we forecast future constraints on the dressed PBH abundance \(f_{\mathrm{PBH}}\) with next-generation radio telescopes such as SKA. Simulating \(10^5\) FRB events under both MMD and log-normal EMD, we find that halo dressing tightens the PBH abundance constraints by roughly an order of magnitude, from \(\sim10^{-3}\) (bare PBHs) to \(\sim10^{-4}\) (dressed PBHs). Moreover, there are two significant aspects in our analysis:
\begin{itemize}
\item When comparing the PBH abundance predicted by a given theoretical model with observational constraints, one must convert the results from the MMD to the EMD within the corresponding framework. For lensing-based constraints, this conversion can be performed using Eq.~(\ref{eq3-19}) and Eq.~(\ref{eq3-20}) for dressed and bare PBHs, respectively. It is important to emphasize that, in the dressed case, the transformation remains unaffected by the evolutionary history of the PBHs; it only alters the effective mass, which in turn changes the strength of the resulting constraints.

\item Gravitational lensing is proposed as a general method to directly identify dressed PBHs and distinguish them from their bare counterparts. As demonstrated in Figs.~\ref{fig:3} and~\ref{fig:4}, microlensing FRBs already provides a sensitive test for such objects when PBHs with masses above the stellar-mass window constitute a sub-percent fraction of dark matter. Upcoming surveys with SKA and other FRB facilities will further explore the relevant parameter space, including regions of interest for gravitational-wave events.
\end{itemize}

Our analysis is subject to several sources of uncertainty. These include, for instance, the theoretical treatment of the dark halo contribution to PBH lensing, which can be estimated by the average convergence and the total Einstein radius \(r_{\mathrm{E,tot}}\) following~\citet{Oguri:2022fir}. In practice, observational determinations of the effective mass $M_{\rm PBH}^{\rm eff}$ require constructing a lens model and extracting characteristic features from the lensed images. Furthermore, additional effects, e.g., clustered PBHs~\cite{DeLuca:2022uvz,Zhang:2025tgm} and the evolution of PBHs from baryonic accretion~\cite{Liu:2023pvq,Zhang:2024ytf}, may also affect the final constraints and should be taken into account in future works.

\section{Acknowledgements}
This work is supported by the Science and Technology Research Project of Hubei Provincial Department of Education under Grants No.Q20251302; National Natural Science Foundation of China under Grants Nos.12322301, 12275021, and 12503002.

\bibliographystyle{apsrev4-1}
\bibliography{ref}
\end{document}